\documentclass[
    ,final            
  ]
  {aipproc}

\layoutstyle{8x11single}



\begin{document}

\title{Dielectric Laser Acceleration}

\author{R. J. England, B. Noble, Z. Wu}{
  address={SLAC National Accelerator Laboratory, Menlo Park, CA 94025, USA}
}

\author{M. Qi}{
  address={Purdue University, West Lafayette, IN 47907, USA}
}

\maketitle



\section{An Investment in Innovation}


{\hskip 0.13in}
 

History has taught us that the greatest advances in accelerator research have come not from incremental improvements to existing technology, but from revolutionary new ideas, often grounded in basic accelerator research conducted years before.  Today, the study of charged particle acceleration is a scientific field in its own right and has produced modern accelerators used in discovery science programs spanning the Office of Science initiative.  Maintaining competitiveness in these programs requires an investment in basic research to identify and develop new accelerator concepts that offer cheaper and more compact alternatives to established techniques, with the goals of improving performance while reducing the size and cost of future colliders and light sources.  Development of such machines will require suitable diagnostics and beam manipulation techniques, including compatible small-footprint deflectors and undulators and high-resolution beam position monitors capable of measuring the overlap of nanometer-scale IP spot sizes.  
\begin{itemize}
\item As a consequence of its unique operating parameter regime, the predicted energy loss due to beam-beam interaction for a dielectric laser accelerator (DLA) based collider is small.  Consequently, among advanced accelerator concepts thus far proposed, \emph{DLA is the only scheme that appears reasonable for a 10 TeV linear collider scenario.}   
 
\item In contrast to other novel accelerator schemes, desirable luminosities would be obtained by operating with very low charge per bunch but at extremely high repetition rates.  Strawman parameters for a 3 TeV scale lepton collider are shown in Table 1.
 
\item The DLA concept leverages well-established industrial fabrication capabilities and the commercial availability of tabletop lasers to reduce cost, while offering significantly higher accelerating gradients, and therefore a smaller footprint.  Power estimates for the DLA scenario are comparable with convention RF technology, assuming that similar power efficiency (near 100\%) for guided wave systems can be achieved.  
 
\item The critical technical challenges for this approach are:  (1) understanding IR laser damage limits of semiconductor materials at picosecond pulse lengths, (2) development of high (near 100\%) efficiency power coupling schemes, (3) integrated designs with multiplee stages of acceleration, and (4) understanding phase stability issues related to temperature and nonlinear high-field effects in dielectrics.  
 
\item Progress towards an energy scalable DLA architecture requires a R\&D focus on fabrication and structure evaluation to optimize existing and proposed concepts, and development of low-charge high-rep rate electron sources that can be used to evaluate performance over many stages of acceleration.
 
\item This research has significant near and long-term applications beyond energy frontier science, including radiation production for compact medical x-ray sources, university-scale free electron lasers, NMR security scanners, and food sterilization.  These additional applications are beginning to be explored.
\end{itemize}

\section{Dielectric Laser Acceleration}

{\hskip 0.13in}
Dielectric laser acceleration refers to the use of infrared (IR) lasers to accelerate charged particles inside of a dielectric waveguide.  The waveguide acts as both the vacuum channel for the beam and as a confining structure to guide an electromagnetic traveling wave mode.  Assuming that the guiding channel's transverse dimensions are of the order of the drive laser wavelength (i.e. 1 to 10 microns) the power coupling efficiency to the particle bunches can in principle be as high as 50\%, with optimal efficiency at bunch charges of 1 to 20 fC.  In order for successive bunches to sit in the accelerating phase of the wave, the requisite bunch durations are on the attosecond scale with intrabunch spacing equal to the laser wavelength (or an integer multiple thereof).  A technique for generating the requisite optically microbunched attosecond scale beams was recently demonstrated at SLAC, and recent work in field emission needle-tip emitters demonstrates that electron beams with the requisite charge and emittance requirements are within reach.   As a result of the various technical requirements just mentioned, the beam parameters for an accelerator based on this technology would be quite different from both traditional machines and other advanced schemes.  

\begin{figure}
 \includegraphics[height=.28\textheight]{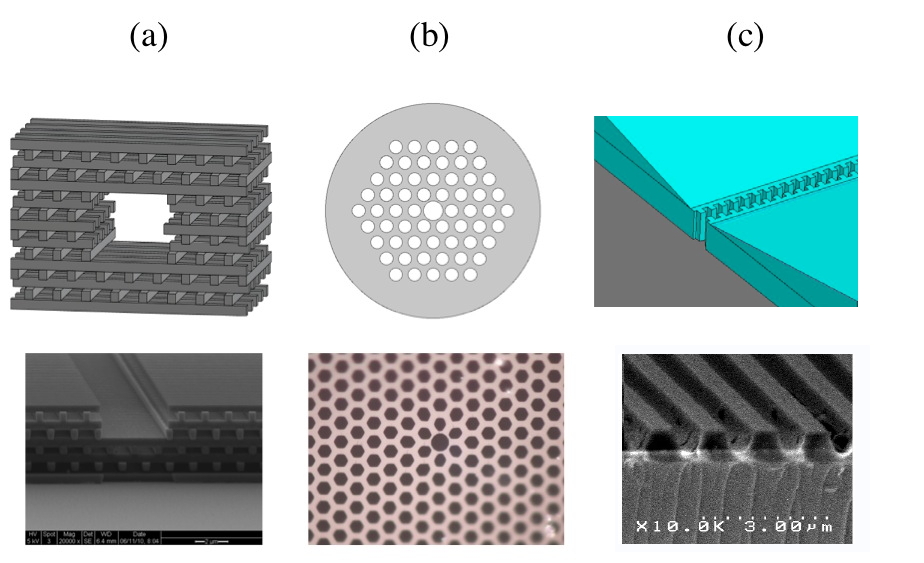}
\caption{Three dielectric laser accelerator topologies:  (a) a 3D silicon photonic crystal structure [courtesy B. Cowan, C. McGuinness], (b) a hollow-core photonic bandgap fiber [courtesy J. Spencer], and (c) a dual-grating structure [courtesy T. Plettner, E. Peralta], showing conceptual illustration (top) and recently fabricated structures (bottom).}
\label{structures}
 \end{figure}

DLA offers several compelling potential advantages over traditional microwave cavity accelerators.  \textbf{Accelerating gradient} is limited by the breakdown threshold for damage of the confining structure in the presence of intense electromagnetic fields.  In the DLA scheme operating at typical laser pulse lengths of 0.1 to 1 ps, the laser damage fluences for dielectric materials such as silicon and glass correspond to peak surface electric fields of 400 to 2000 MV/m.  This is to be compared with breakdown limits of 40 to 100 MV/m for metal cavities.  The corresponding gradient enhancement represents a reduction in active length of the accelerator between 1 and 2 orders of magnitude.  \textbf{Power sources} for DLA-based accelerators (lasers) are cheaper than microwave sources (klystrons) for equivalent average power levels due to the wider availability and private sector investment in commercial laser sources.  Due to the high laser-to-particle coupling efficiency, required pulse energies are consistent with tabletop microjoule class lasers.  \textbf{Fabrication techniques} for constructing three-dimensional dielectric structures with nanometer-level precision are well established in the semiconductor industry and the capillary fiber industry.  Once a suitable fabrication recipe is developed, on-chip DLA devices with multiple stages of acceleration and waveguides for coupling power to and from the structure could be manufactured at low per-unit cost on silicon wafers.

Several DLA topologies are under investigation, as seen in Fig. \ref{structures}:  (a) a silicon "woodpile" photonic crystal waveguide, (b) a glass photonic bandgap (PBG) hollow-core optical fiber, and (c) a structure where the beam is accelerated by a transversely incident laser beam in the gap between two gratings.  Significant progress has been made in the fabrication of partial or full prototypes of these structures with geometries optimized for accelerator use, as seen in the bottom images.  One of the first completed DLA prototype structures, recently fabricated at the Stanford Nanofabrication Facility, has been tested at the Next Linear Collider Test Accelerator at SLAC, as shown in Fig. 2(a).  Accelerated electron energies of 26 keV over a 0.5mm interaction length were observed.  These results are the first clear indication of particle acceleration within an enclosed laser-driven dielectric structure operating at optical wavelengths, and already showing promising accelerating gradients.

 \begin{figure}
 \includegraphics[height=.28\textheight]{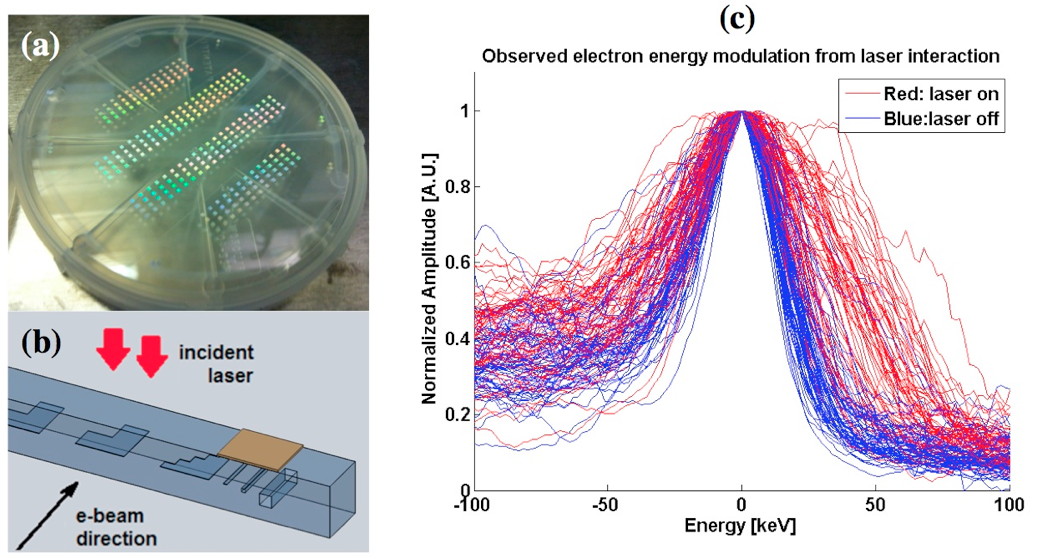}
\caption{Image (a) of fully bonded wafers containing multiple DLA prototypes and (b) inset schematic of a single diced test sample cut from this wafter, and (c) transmitted electron spectrum showing energy broadening of the electrons for laser on (red) vs. laser off (blue) events.}
\label{dualgrating}
 \end{figure}

\section{LINEAR COLLIDER STRAWMAN}

{\hskip 0.13in}
To reach 10 TeV center-of-mass energies, a next generation lepton collider based on traditional RF microwave technology would need to be over 100 km in length and would likely cost tens of billions of dollars to build.  Due to the inverse scaling of the interaction cross section with energy, the required luminosity for such a machine would be as much as $100\times$ higher than proposed 1-3TeV machines (ILC and CLIC), producing a luminosity goal of order $10^{36}$ cm$^{-2}$s$^{-1}$.  In attempting to meet these requirements in a smaller cost/size footprint using advanced acceleration schemes, the increased beam energy spread from radiative loss during beam-beam interaction (beamstrahlung) at the interaction point becomes a pressing concern.  Since the beamstrahlung parameter is proportional to bunch charge, a straightforward approach to reducing it is to use small bunch charges, with the resulting quadratic decrease in luminosity compensated by higher repetition rates.  This is the natural operating regime of the DLA scheme, with the requisite average laser power (>100 MW) and high (>10 MHz) repetition rates to be provided by modern fiber lasers.

In Table \ref{tbl_lc}, we compare strawman parameters for a DLA based collider and Higgs factory with expected parameters for CLIC at 3TeV.  In these examples, DLA meets the desired luminosity, and with a significantly smaller beamstrahlung energy loss.  Other advanced collider schemes such as beam-driven plasma and terahertz also rely upon a traditional pulse format for the electron/positron beam and would therefore compare similarly to CLIC in this regard.  Although the numbers in Table \ref{tbl_lc} are merely projections used for illustrative purposes, they highlight the fact that due to its unique operating regime, DLA is poised as a promising technology for future collider applications.  Calculations in Table 1 for the DLA cases assume the full bunch train charge per crossing.  This assumption is valid for the geometrical luminosity if the beta function at the interaction point is larger than the bunch train length (true for most configurations of interest).  It is valid for the beamstrahlung and luminosity enhancement (here a factor of 6) if the microbunching is undone after acceleration but prior to the IP, such as by introducing a dispersive section to smear out the longitudinal modulation.  

 
 
 \begin{table}[htb] 
\caption{Strawman Parameters for DLA Linear Collider.} \label{tab:GoverningRelations}
\label{tbl_lc}
\centering \begin{tabular}{| c||c | c | c | c |}
\hline
{\bf Parameter}& {\bf Units} & {\bf CLIC}& {\bf DLA 3TeV} & {\bf DLA 250 GeV} \\ \hline
Center-of-Mass Energy        &   GeV                & 3000  &            3000             &  250        \\
Bunch Charge             &   e           &  3.7E+09            &            38000              &  38000 \\ 
Bunches per Train &  & 312 & 159 & 159 \\
Train Repetition Rate  &  MHz   &   5.0E-5  &  30    &   60   \\
Bunch Train Length  &   ps   &   26005   &    1.0   &   1.0   \\
Single Bunch Length   &   $\mu$m   &   34.7   &   0.0026   &   0.0026   \\
 & & & &  \\
Design Wavelength     &   $\mu$m   &  230609   &   2.0    &   2.0 \\
Invariant X Emittance   &   $\mu$m   &  0.66   &   0.0001   &   0.002   \\
Invariant Y Emittance   &   $\mu$m   &  0.02   &   0.0001   &   0.002   \\
IP X Spot Size   &   nm   &   45   &   1   &   2   \\
IP Y Spot Size   &   nm   &   1   &   1   &   2   \\
Beamstrahlung Energy Loss  &   \%   &    36.2    &    1.1    &   0.6    \\
\bf Enhanced Luminosity / top 1\%  & \bf cm$^{-2}$/s  & \bf  8.6E+34   & \bf  8.1E+34   & \bf 1.3E+34  \\
& & & & \\
Beam Power  &  MW  &  13.9   &   43.6   &   7.3  \\
& & & & \\
Wall-Plug Efficiency  &  \%  &  7.0  &  15.1  &  15.1  \\
Wall-Plug Power  &  MW  &  200  &  289  &  48  \\
Gradient  &  MV/m  &  100  &  400  &  400  \\
Total Linac Length  &  km  &  42.0  &  7.5  &  0.6  \\
\hline
\end{tabular} 
\end{table}

\begin{figure}
 \includegraphics[height=.30\textheight]{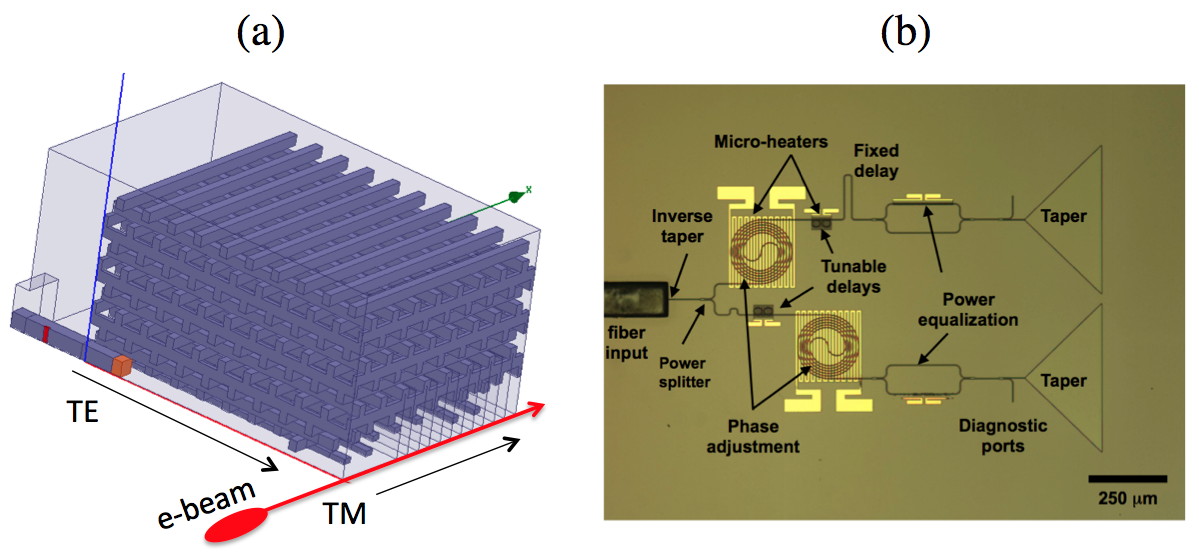}
\caption{Images of (a) HFSS simulation geometry for high-efficiency transverse power coupling to the woodpile structure and (b) recently prototyped SiN waveguide network for power distribution from a single laser feed.}
\label{woodpile_coupling}
 \end{figure}

Obtaining wall-plug efficiencies suitable for linear collider applications will additionally require the development of integrated couplers with high efficiency, fed by a network of waveguides that split the laser power from a common feed among various accelerator components. Initial results in simulating such couplers, shown in Fig. \ref{woodpile_coupling}(a) for the woodpile structure using silicon-on-insulator (SOI) waveguides indicate coupling efficiencies from the input waveguide to the accelerating mode close to 100\% \cite{Wu:PAC11}.  The power distribution scheme is then envisioned as a fiber-to-SOI coupler that brings a pulse from an external fiber laser onto the integrated chip, distributes it between multiple structures via SOI power splitters, and then recombines the spent laser pulse and extracts it from the chip via a mirror-image SOI-to-fiber output coupler \cite{Colby:PAC11}, after which the power is either dumped, or for optimal efficiency, recycled.  For the DLA calculations in Table \ref{tbl_lc}, accelerator to electron coupling efficiencies of 40\% are assumed, consistent with the estimates of Ref. \cite{Siemann:PRSTB04}.  Assumed laser wall-plug efficiency is 40\%, feasible using current solid state Thulium-doped fiber laser technology, which has already achieved wall-plug efficiencies above 30\% \cite{Moulton:2009}.  Maintaining phase synchronicity of the laser pulse and the accelerated electrons between many separately fed structures would be accomplished by fabricating the requisite phase delays into the lengths of the waveguide feeds.  Prototype grating to SOI couplers have recently been produced for 1.5 $\mu$m wavelength operation, images of which are shown in Fig. \ref{woodpile_coupling}(b).  
 
\section{Additional Applications}
{\hskip 0.13in}

Realizing the promise of DLA structures to provide low cost compact accelerators for a wide variety of uses requires that other compatible accelerator components be developed, several of which have exciting potential for both near and longer term applications.  \textbf{A dielectric laser-driven deflector} was recently proposed by Plettner and Byer \cite{Plettner:PRSTAB08}, that uses a pair of dielectric gratings excited transversely by a laser beam and separated by a gap of order the laser wavelength where a beam of electrons would travel.  By changing the sign of the excitation between successive structures (e.g. by alternating the direction of illumination) an optically powered undulator could be constructed to create laser driven micro-undulators for production of \textbf{attosecond-scale radiation pulses} synchronized with the electron bunch.  Unlike other electromagnetic undulator concepts, the undulator period in this scheme is set by the length of each deflection stage and can therefore be much larger than the driving wavelength.  Undulators based upon this concept could attain very short (mm to sub-mm) periods with multi-Tesla field strengths: an undulator with a 250 $\mu$m period driven by a 2 $\mu$m solid state laser would have a gain length of 4 cm and an X-ray photon energy of 10 keV when driven by a 500 MeV electron beam.  Since DLA structures operate optimally with optical-scale electron bunch formats, high repetition rate (10s of MHz) attosecond-scale pulses are a natural combination.  

The grating-based optical deflector concept presents the possibility of creating a FEL undulator compatible with future laser-driven accelerators.  We have simulated a single period of such an undulator concept using an electromagnetic finite element code with a grating excited transversely by a Gaussian monochromatic laser mode.  The axial fields extracted show not only a significant integrated transverse deflecting force, but also evidence of harmonic field components.  Since these harmonics can be manipulated simply by altering the transverse profile of the laser beam (unlike a traditional magnetic undulator where the field profile is fixed), this could provide a flexible alternative to harmonic seeding schemes designed to extend the range of wavelengths accessible via coupling to higher harmonics of the fundamental FEL radiation process.

\section{Meeting Future Challenges Through Collaboration and Education}

Dielectric laser acceleration is a multidisciplinary field, drawing upon expertise in infrared laser technology, materials science, beam dynamics, semiconductor fabrication techniques, and experimental accelerator physics.  Following initial proof-of-principle demonstrations of dielectric laser acceleration, which are anticipated to take place within the next year, the challenge will be to develop this technique into a useful acceleration method. Among the issues that need to be resolved are:  (1) understanding IR laser damage limits of semiconductor materials at picosecond pulse lengths; (2) development of high (near 100\%) efficiency schemes for coupling fiber or free space lasers into DLA structures; (3) developing integrated designs with multiple stages of acceleration; and (4) understanding phase stability issues related to temperature and nonlinear high-field effects in dielectrics.

Answering these questions will require strong collaborations between university and laboratory groups that can attract some of the brightest researchers in the field of beam physics.  In the last 10 years, the Advanced Accelerator Research Department at SLAC has hosted over 70 PhD students, of whom nearly half were from other institutions, and nearly 40 postdocs, with alumni proceeding to successful careers in industry, university, and national laboratory positions.  Appropriate investment in these facilities will allow the continued development of DLA and other advanced acceleration concepts, and will provide opportunities for students, post-docs, and scientists from SLAC, Stanford and other institutions across the world to engage in ground-breaking experimental, theoretical, and computational work at the cutting edge of accelerator research.



\bibliographystyle{aipproc}   

\bibliography{england}

\IfFileExists{\jobname.bbl}{}
 {\typeout{}
  \typeout{******************************************}
  \typeout{** Please run "bibtex \jobname" to optain}
  \typeout{** the bibliography and then re-run LaTeX}
  \typeout{** twice to fix the references!}
  \typeout{******************************************}
  \typeout{}
 }

\end{document}